\documentclass[10pt,conference]{IEEEtran}
\usepackage{graphicx}
\usepackage{amsmath}
\usepackage{amssymb}
\usepackage{theorem}

\newif\ifambproofs
\newif\ifdraft
\newif\iffigures

\graphicspath{{./}{/home/jungp/TeX/Figures/}}

\theoremstyle{plain}
\newtheorem{mytheorem}{Theorem}

\newtheorem{mylemma}[mytheorem]{Lemma}

\theoremstyle{definition}

\newtheorem{mydefinition}[mytheorem]{Definition}

\newenvironment{myproof}{{\it Proof:$\quad$}}{$\quad$\QED \\}

\usepackage{graphics,graphicx,color}

\newcommand{\WLOG}{w.l.o.g. }

\newcommand{\sympl}{\eta}

\newcommand{\Trace}[1]{\,\mathbf{Tr}{#1}}
\newcommand{\setZ}{{\mathbb{Z}}}

\newcommand{\schattenclass}{\mathcal{T}}
\newcommand{\settraceclass}{\schattenclass_1}

\newcommand{\defeq}{\overset{\text{def}}{=}}

\newcommand{\HH}{{\mathcal{H}}}
\newcommand{\BH}{\boldsymbol{\HH}}

\newcommand{\BHSpread}{\boldsymbol{\Sigma}}
\newcommand{\BHWeyl}{\boldsymbol{L}}

\newcommand{\Leb}[1]{\mathcal{L}_{#1}}
\newcommand{\Ltwo}{\Leb{2}}

\newcommand{\Amb}{{\mathbf{A}}}

\newcommand{\Real}[1]{{\text{Re}}\{#1\}}

\newcommand{\Shift}{{\boldsymbol{S}}}

\newcommand{\sFourier}{\mathcal{F}_s}
\newcommand{\nFourier}{\mathbf{F}}

\newcommand{\esup}{\text{\rm ess sup\,\,}}

\newcommand{\Reals}{\mathbb{R}}
\newcommand{\RealsPlus}{\mathbb{R}_{+}}
\newcommand{\Complexes}{\mathbb{C}}

\begin{document}
\title{
  Approximate Eigenstructure of LTV Channels 
  with Compactly Supported Spreading
}
\author{Peter Jung\\
  Fraunhofer German-Sino Lab for Mobile Communications (MCI) and the Heinrich-Hertz Institute, Berlin\\[.1em]
  \small{jung@hhi.fraunhofer.de}}
\maketitle
\begin{abstract}
   In this article we obtain estimates on
   the approximate eigenstructure of channels with 
   a spreading function supported only on a set of finite measure $|U|$.
   Because in typical application like wireless communication
   the spreading function is a random process corresponding to a random
   Hilbert--Schmidt channel operator $\BH$
   we measure this approximation in terms of the ratio of 
   the $p$--norm of the deviation from variants of the Weyl symbol calculus to
   the $a$--norm of the spreading function itself. This generalizes recent results
   obtained for the case $p=2$ and $a=1$. We provide a general approach to this
   topic and consider then operators with $|U|<\infty$
   in more detail. We show the relation to pulse shaping and weighted norms
   of ambiguity functions. Finally we derive several necessary conditions
   on $|U|$, such that the approximation error is below certain levels.
\end{abstract}

%%%%%%%%%%%%%%%%%%%%%%%%%%%%%%%%%%%%%%%%%%%%%%%%%%%%%%%%%%%%%%%%%%%%%%%%%
%%%%%%%%%%%%%%%%%%%%%%%%%%%%%%%%%%%%%%%%%%%%%%%%%%%%%%%%%%%%%%%%%%%%%%%%%
%
\section{Introduction}
Optimal signaling through linear time--varying (LTV) channels is a challenging
task for future communication systems. For a particular realization 
of the time--varying channel operator the transmitter and receiver
design which avoids crosstalk between different time--frequency slots
is related to ''eigen--signaling''. Eigen--signaling simplifies much the
information theoretic treatment of communication in dispersive channels.
However, it is well--known that
for a whole class of channels such a joint separation of the 
subchannels can not be achieved. A typical scenario, present for example
in wireless communication, is signaling through a random doubly--dispersive
channel $\BH$:
\begin{equation*}
   r(t)=(\BH s)(t)+n(t)
\end{equation*}
From signal processing point of view 
the preferred design of the transmit signal $s(t)$ needs knowledge
on the true eigenstructure of $\BH$. This would in principle
allow interference--free transmission and simple recovering algorithms of the information
from received signal $r(t)$ degraded by the noise process $n(t)$. 
However, for $\BH$ being random, random eigenstructure has to be 
expected in general and a joint design of the transmitter and the receiver
for an ensemble of channels has to be performed. Nevertheless with 
such an approach interference can not be avoided and
remains in the communication chain. For such interference scenarios it
is important to have bounds on the distortion of a particular selected 
signaling scheme.

First results in this field can be found 
already in the literature on pseudo--differential operators 
\cite{kohn:pdo:algebra,folland:harmonics:phasespace}.
More recent results with direct application to time--varying channels
were obtained by Kozek \cite{kozek:thesis}
and Matz \cite{matz:thesis} which resemble the notion of underspread
channels. They investigated the approximate symbol calculus of pseudo--differential
operators in this context and derived bounds for the $\ell_2$--norm
of the distortion which follow from the approximate product rule 
in terms of Weyl symbols. Controlling this approximation
intimately scales with the ''size'' of the spreading of the contributing 
channel operators. For operators with compactly supported spreading
this is $|U|$ -- the size of the spreading support $U$.
Interestingly this approximation behavior breaks 
down in their framework at a certain critical size. Channels 
below this critical size are called in their terminology underspread, 
otherwise overspread.
  
Underspreadness of time--varying channels occurs also in the context
of channel measurement \cite{kailath:ltvmeasurement}. See
also the recent article \cite{kozek:identification:bandlimited} for a 
rigorous treatment of channel identification
based on Gabor (Weyl--Heisenberg) frame theory. The authors connect
the critical time--frequency sampling density immanent in this theory to
the stability of the channel measurement. A relation between these different
notions of underspreadness has to be expected but will be out of 
the scope of this paper.

This article considers the problem of approximate eigenstructure
from a different angle, namely investigating the $\ell_p$--norm $E_p$ 
of the error $\BH s-\lambda r$
for well--known choices of $\lambda$.
This direct formulation allows for improvements to the existing bounds,
generalizations to arbitrary regions of spreading and 
different distortion measures.  Furthermore the approach 
will show the connection to well--known fidelity criteria related to pulse design 
\cite{kozek:thesis,jung:wssuspulseshaping,jung:isit06}. 
Using the techniques recently 
presented in \cite{jung:isit06} we finally extract necessary conditions for $|U|$ 
that the error does not exceed certain levels. Furthermore
we discover some interesting relations related to underspreadness in form of 
\cite{kozek:thesis,matz:thesis}.

The paper is organized as follows. After settling the basic definitions in
the first section of the paper the second section 
reviews the Weyl correspondence and the
spreading representation of Hilbert--Schmidt operators. 
In the next sections we will then consider the problem of controlling $E_p$
and give necessary conditions on $|U|$. 
In the last part we will verify our framework with some numerical tests.

%%%%%%%%%%%%%%%%%%%%%%%%%%%%%%%%%%%%%%%%%%%%%%%%%%%%%%%%%%%%%%%%%%%%%%%%%
%%%%%%%%%%%%%%%%%%%%%%%%%%%%%%%%%%%%%%%%%%%%%%%%%%%%%%%%%%%%%%%%%%%%%%%%%
\subsection{Some Definitions}
But before starting, the following definitions are needed.
For $1\leq p<\infty$ and $f:\Reals\rightarrow\Complexes$  
the functional $\lVert f\rVert_p\defeq\left(\int|f(t)|^pdt\right)^{1/p}$
is then usual notion of the $p$--norm
($dt$ is the Lebesgue measure on $\Reals$).
Furthermore for $p=\infty$ is
$\lVert f\rVert_\infty\defeq\esup |f(t)|$.
If $\lVert f\rVert_p$ is finite $f$ is said to be in $\Leb{p}(\Reals)$. 
We will frequently make use of the relation
$\lVert f^a\rVert_b^c=\lVert f\rVert^{ab}_{ac}$. The function
$\chi_U$ will always denote the characteristic function onto the set $U\subseteq\Reals^2$.

%%%%%%%%%%%%%%%%%%%%%%%%%%%%%%%%%%%%%%%%%%%%%%%%%%%%%%%%%%%%%%%%%%%%%%%%%
\section{Displacements Operators and Ambiguity Functions}
%%%%%%%%%%%%%%%%%%%%%%%%%%%%%%%%%%%%%%%%%%%%%%%%%%%%%%%%%%%%%%%%%%%%%%%%%
Time-frequency representations are an important tool in signal analysis and physics. 
Among them are Woodward's cross ambiguity function %\cite{woodward:probinfradar}
and the Wigner distribution. Ambiguity functions can be understood as 
inner products representations of
displacement (or shift) operators,
defined by its action on function $f:\Reals\rightarrow\Complexes$ as:
\begin{equation}
   (\Shift_{\mu} f)(x):=e^{i2\pi\mu_2x}f(x-\mu_1)
   \label{eq:shiftop}
\end{equation}
The functions $f$ are time--signals, i.e. $\mu=(\mu_1,\mu_2)\in\Reals^2$ where
$\mu_1$ is time shift and $\mu_2$ is a frequency shift.
However the following can be straightforward extended to multiple dimensions. 
The operator $\Shift_\mu$ acts isometrically on all $\Leb{p}(\Reals)$, hence is unitary 
on the Hilbert space $\Leb{2}(\Reals)$. 
The arbitrariness due to the non--commutativity of shifts in the previous definition 
can be covered in using generalized displacements of
the form: 
\begin{equation}
   \begin{split}
      \Shift_{\mu}(\alpha)
      &=
      \Shift_{(0,\mu_2(\frac{1}{2}+\alpha))}
      \Shift_{(\mu_1,0)}
      \Shift_{(0,\mu_2(\frac{1}{2}-\alpha))}
   \end{split}
   \label{eq:weyl:shift:alphageneralized}
\end{equation}
We will call $\alpha$ as polarization.
All these operators establish (up to unitary equivalence) 
unitary representations of the Weyl--Heisenberg
group on $\Ltwo(\Reals)$ (see for example \cite{folland:harmonics:phasespace}). 
In physics it is common to choose the most symmetric case $\alpha=0$ 
and the operators are usually  called  Weyl operators.
The definition \eqref{eq:shiftop}
appears for $\alpha=1/2$ and we call
$\Shift_\mu=\Shift_\mu(1/2)$ as
\emph{time--frequency shift operator}.
If we define the symplectic form as
$\sympl(\mu,\nu):=\mu_1\nu_2-\mu_2\nu_1$,
we have the following well--known 
\emph{Weyl commutation relation}:
\begin{equation}
   \begin{split} 
      \Shift_{\mu}(\alpha)\Shift_{\nu}(\beta) =
      e^{-i2\pi \sympl(\mu,\nu)}\Shift_{\nu}(\beta)\Shift_{\mu}(\alpha)
   \end{split} 
   \label{eq:weyl:shift:comm}
\end{equation}
In this way the \emph{cross ambiguity function} can be defined as:
$\Amb^{(\alpha)}_{g\gamma}(\mu)\defeq\langle g,\Shift_\mu(\alpha)\gamma\rangle$.

%%%%%%%%%%%%%%%%%%%%%%%%%%%%%%%%%%%%%%%%%%%%%%%%%%%%%%%%%%%%%%%%%%%%%%%%%
\section{Weyl Correspondence and the Spreading Representation}
%%%%%%%%%%%%%%%%%%%%%%%%%%%%%%%%%%%%%%%%%%%%%%%%%%%%%%%%%%%%%%%%%%%%%%%%%

In the previous section it was motivated that
$\Amb^{(\alpha)}_{g\gamma}$ yields a local time--frequency description of functions. 
Now the same can be repeated for Hilbert--Schmidt operators. 
Let us introduce them as the $2$th Schatten class:
If we define for linear mappings $A\in L(\Leb{2}(\Reals))$ from 
Hilbert space $\Leb{2}(\Reals)$
into itself $|A|:=(A^*A)^{1/2}$, then for $1\leq p<\infty$
the functional 
$\lVert A\rVert_p:=\Trace{(|A|^p)}^{1/p}$ 
is 
called the $p$th Schatten norm. The set 
$\schattenclass_p:=\{A\in L(\Leb{2}(\Reals)),\lVert A\rVert_p<\infty\}$
is called the $p$th Schatten class where 
$\schattenclass_\infty$ is set to be the compact operators.
Then $\schattenclass_p$ for $1\leq p<\infty$ are Banach spaces and 
$\schattenclass_1\subset\schattenclass_p\subset\schattenclass_\infty$ 
(see for example \cite{reed:simon:fourier} or \cite{holevo:propquantum}). 
The sets $\settraceclass$ and $\schattenclass_2$ 
are called trace class and Hilbert--Schmidt operators. Hilbert--Schmidt operators
form itself a Hilbert space with inner product
$\langle A,B\rangle_{\schattenclass_2}\defeq\Trace{A^*B}$.

In particular for any $\BH\in\settraceclass$ there holds by
properties of the trace $\Trace(X\BH)\leq\lVert \BH\rVert_1\lVert X\rVert$, where
$\lVert\cdot\rVert$ denotes the operator norm.
Hence for $X=\Shift_\mu(\alpha)$ given by \eqref{eq:weyl:shift:alphageneralized} one can 
define with analogy to ordinary Fourier transform \cite{daubechies:integraltransform,grossmann:wignerweyliso}
a mapping $\nFourier^{(\alpha)}:\settraceclass\rightarrow\Leb{2}(\Reals^{2})$ via
\begin{equation}
   (\nFourier^{(\alpha)} \BH)(\mu)\defeq
   \Trace(\Shift_\mu^*(\alpha)\BH)=\langle\Shift_\mu(\alpha),\BH\rangle_{\schattenclass_2}
   \label{eq:tfanalysis:noncomm:fourier}
\end{equation}
Note that $(\nFourier^{(\alpha)} \BH)(0)=\Trace \BH$ and 
$|(\nFourier^{(\alpha)} \BH)(\mu)|\leq\lVert \BH\rVert_1$. 
The function
$\nFourier^{(\alpha)} \BH\in\Leb{2}(\Reals^{2})$ is sometimes
called the ''non--commutative'' Fourier transform \cite{holevo:propquantum}, 
inverse Weyl transform \cite{weyl:theoryofgroups:quantum} 
or $\alpha$--generalized \emph{spreading function} of $\BH$ \cite{kozek:thesis}.
\begin{mylemma}[Spreading Representation]
   Let $\BH\in\schattenclass_2$. Then there holds
   \begin{equation}
      \BH
      =\int (\nFourier^{(\alpha)} \BH)(\mu)\Shift_{\mu}(\alpha) d\mu
      =\int \langle\Shift_\mu(\alpha),\BH\rangle_{\schattenclass_2}\Shift_{\mu}(\alpha) d\mu
      \label{eq:tfanalysis:linop:spreading}
   \end{equation}
   where the integral is meant in the weak sense. 
   \label{lemma:tfanalysis:linop:spreading}
\end{mylemma}
The extension  to $\schattenclass_2$ is
due to continuity of 
$\nFourier^{(\alpha)}:\settraceclass\rightarrow\Leb{2}(\Reals^{2})$ and density of 
$\schattenclass_1$ in $\schattenclass_2$.
A complete proof of this lemma can be found for example in \cite{holevo:propquantum}.
For $\BH,X\in\schattenclass_2$ the following Parseval-like identity
\begin{equation}
   \langle X,\BH\rangle_{\schattenclass_2}=
   \langle \nFourier^{(\alpha)} X,\nFourier^{(\alpha)}\BH\rangle
   \label{eq:tfanalysis:spreading:parceval}
\end{equation}
holds.
If we define the symplectic Fourier transform of a function 
$\sFourier:\Reals^2\rightarrow\Complexes$ as: 
\begin{equation}
   (\sFourier F)(\mu)=\int_{\Reals^2} e^{-i2\pi\sympl(\nu,\mu)} F(\nu)d\nu
   \label{eq:tfanalysis:symplectic:fourier}
\end{equation}
then $\sFourier\nFourier^{(\alpha)}$ establishes a correspondence between 
the ordinary function 
$\BHWeyl^{(\alpha)}_{X}=\sFourier\nFourier^{(\alpha)}X$
and an operator $X$ (Weyl quantization \cite{weyl:theoryofgroups:quantum}).
The function $\BHWeyl^{(\alpha)}_{X}$ is called (generalized) \emph{Weyl symbol} of $X$.
The original Weyl symbol is $\BHWeyl^{(0)}_{X}$. The cases $\alpha=1/2$ and
$\alpha=-1/2$ are also known as Kohn--Nirenberg symbol (or Zadeh's 
time--varying transfer function) and Bello's 
frequency--dependent modulation function \cite{bello:wssus}. 
Using Parseval identity for $\sFourier$ eq. \eqref{eq:tfanalysis:spreading:parceval}
extends now to
\begin{equation}
   \langle X,Y\rangle_{\schattenclass_2}=
   \langle \nFourier^{(\alpha)} X,\nFourier^{(\alpha)}Y\rangle=
   \langle \BHWeyl_X^{(\alpha)},\BHWeyl_Y^{(\alpha)}\rangle
   \label{eq:tfanalysis:weyl:parceval}
\end{equation}
and consequentially 
$\lVert X\rVert_2=\lVert\nFourier^{(\alpha)}X\rVert_2=\lVert\BHWeyl_X^{(\alpha)}\rVert_2$.
%%%%%%%%%%%%%%%%%%%%%%%%%%%%%%%%%%%%%%%%%%%%%%%%%%%%%%%%%%%%%%%%%%%%%%%%%
%%%%%%%%%%%%%%%%%%%%%%%%%%%%%%%%%%%%%%%%%%%%%%%%%%%%%%%%%%%%%%%%%%%%%%%%%
%
\section{Eigenstructure of Operators with Compactly Supported Spreading}
%
%%%%%%%%%%%%%%%%%%%%%%%%%%%%%%%%%%%%%%%%%%%%%%%%%%%%%%%%%%%%%%%%%%%%%%%%%
%%%%%%%%%%%%%%%%%%%%%%%%%%%%%%%%%%%%%%%%%%%%%%%%%%%%%%%%%%%%%%%%%%%%%%%%%

%%%%%%%%%%%%%%%%%%%%%%%%%%%%%%%%%%%%%%%%%%%%%%%%%%%%%%%%%%%%%%%%%%%%%%%%%
\subsection{The Approximate Eigenstructure}
%%%%%%%%%%%%%%%%%%%%%%%%%%%%%%%%%%%%%%%%%%%%%%%%%%%%%%%%%%%%%%%%%%%%%%%%%
It is of general importance how much the Weyl symbol or a smoothed version 
of it approaches the 
eigenvalue characteristics of a given Hilbert--Schmidt operator. 
Since $\BH\in\schattenclass_2$ is compact, it 
has a Schmidt representation\footnote{For $\BH$ given in matrix representation
  also known as ''Singular Value Decomposition''}
$\BH=\sum_{n=1}^\infty s_k\langle x_k,\cdot\rangle y_k$
with the singular values $\{s_k\}$ of $\BH$ and orthonormal bases
$\{x_k\}$ and $\{y_k\}$. However,
the latter depends explicitely on $\BH$ and can be very unstructured.
We are interested in a choice which is ''more independent'' of $\BH$, which
give rise to the following definition of what we will
call ''approximative'' eigenstructure:
\begin{mydefinition}[Approximate Eigenstructure]
   Let $\lambda\in\Complexes$. If for  
   $g,\gamma:\Reals\rightarrow\Complexes$ 
   with $\lVert g\rVert_p=\lVert \gamma\rVert_p=1$ holds
   \begin{equation}
      \lVert \BH\gamma-\lambda g\rVert_p\leq\epsilon(\lambda,\gamma,g)
      \label{eq:approxeigen1}
   \end{equation}
   we call $\lambda$ an ''$\ell_p$--approximate eigenvalue''  of $\BH$ with 
   bound $\epsilon(\lambda,\gamma,g)$.
\end{mydefinition}
Obviously we have $\epsilon(s_k,x_k,y_k)=0$ for each $p$. 
The question is, how much particular choices for
$\lambda(\mu)$ and 
functions $\Shift_\mu\gamma$ and $\Shift_\mu g$ 
can approach the eigenstructure of a Hilbert Schmidt operator $\BH$.
Because in general these functions differ from the Schmidt representation of $\BH$ they
will give an error which we have to control.
Summarizing\footnote{Note that 
  $\Shift_\mu$ can be replaced with $\Shift_\mu(\beta)$ without change of $E_p$}:
\begin{equation}
   \begin{split}
      E_p:
      &=\lVert\BH\Shift_\mu\gamma-\lambda(\mu)\Shift_\mu g\rVert_p
      \leq \epsilon(\lambda(\mu),\Shift_\mu\gamma,\Shift_\mu g)
   \end{split}
   \label{eq:approxeigen:epweyl}
\end{equation}
At this point we introduce furthermore the following abbreviation:
$\BHSpread_{\BH}^{(\alpha)}:=\nFourier^{(\alpha)}\BH$ will always
be the spreading function of $\BH$. With $U\subseteq\Reals^2$ 
we will denote its support and with $|U|=\lVert\chi_U\rVert_1$ 
its size (measure).
\subsection{Previous Results}
In \cite[Theorem 5.6]{kozek:thesis} W. Kozek has been considered the case
$\lambda=\BHWeyl^{(0)}_{\BH}=\sFourier\BHSpread^{(0)}_{\BH}$ and $g=\gamma$.
He obtained the following result:
\begin{mytheorem}[W. Kozek \cite{kozek:thesis}]
   Let $U=[-\tau_0,\tau_0]\times[-\nu_0,\nu_0]$. If $|U|=4\tau_0\nu_0\leq 1$
   then
   \begin{equation}
      E_2^2\leq
      2\sin(\frac{\pi |U|}{4})\lVert\BHSpread^{(0)}_{\BH}\rVert^2_1+
      \epsilon_\gamma\left(\lVert\BHSpread^{(0)}_{\BH^*\BH}\rVert_1+
      2\lVert\BHSpread^{(0)}_{\BH}\rVert^2_1\right)
    \label{eq:approxeigen:kozektheorem}
   \end{equation}
   where $\epsilon_\gamma=\lVert(\Amb^{(0)}_{\gamma\gamma}-1)\chi_U\rVert_\infty$.
   \label{thm:approxeigen:kozektheorem}
\end{mytheorem}
Using $\lVert\BHSpread^{(0)}_{\BH^*\BH}\rVert_1\leq\lVert\BHSpread^{(0)}_{\BH}\rVert^2_1$ 
eq. \eqref{eq:approxeigen:kozektheorem} can be written as
\begin{equation}
   \frac{E_2^2}{\lVert\BHSpread^{(0)}_{\BH}\rVert^2_1}
   \leq 2\sin(\frac{\pi |U|}{4}) + 3\lVert(\Amb^{(0)}_{\gamma\gamma}-1)\chi_U\rVert_\infty
   \label{eq:approxeigen:kozektheorem:simplified}
\end{equation}
G. Matz generalized the result of Theorem \ref{thm:approxeigen:kozektheorem} 
in \cite[Theorem 2.22]{matz:thesis} 
to a formulation in terms of weighted $1$--moments of
spreading functions which
includes now different polarizations $\alpha$ and is not restricted
to the special choice of $U$. For
$U=[-\tau_0,\tau_0]\times[-\nu_0,\nu_0]$ and $\alpha=0$ the bounds
agree with \eqref{eq:approxeigen:kozektheorem:simplified}.
%Both results could be interpreted in such a way that only the second term can controlled
%by $\gamma$ (pulse shaping) where the first of the rhs constant term
%related only the overall spread $|U|$.
\subsection{New Related Results}
\newcommand{\brho}{\bar{\rho}}
\newcommand{\brhoi}{{\bar{\rho}_\infty}}
Instead of directly considering the case of using the Weyl symbol for 
the approximation of the eigenstructure, we use a ''smoothed'' version:
\begin{equation}
   \lambda=\sFourier (\BHSpread^{(\alpha)}_{\BH}\cdot B)
   \label{eq:approxeigen:lambda}
\end{equation}
and consider two cases:
\begin{itemize}
\item[{\bf C1:}] ''$B=\Amb^{(\alpha)}_{g\gamma}$'', such that    
   $\lambda=\BHWeyl_{\BH}^{(\alpha)}\ast\sFourier\Amb^{(\alpha)}_{g\gamma}$ where 
   $\ast$ denotes convolution. This corresponds to the well--known 
   smoothing with the cross Wigner function $\sFourier\Amb^{(\alpha)}_{g\gamma}$.
\item[{\bf C2:}] ''$B=1$'', such that $\lambda(\mu)=\BHWeyl_{\BH}^{(\alpha)}(\mu)$. This case
   is related to the symbol calculus and needed 
   for comparisons with the previous results given so far.
\end{itemize}

Then the following theorem parallels Theorem \ref{thm:approxeigen:kozektheorem} 
and its consequence
\eqref{eq:approxeigen:kozektheorem:simplified}.
\begin{mytheorem}
   For $1\leq p<\infty$ and $1\leq a\leq\infty$ holds:
   \begin{equation}
      \frac{E^p_p}{\lVert\BHSpread^{(\alpha)}_{\BH}\rVert^p_a}\leq
      \brhoi^{p-2}\lVert(1+|B|^2-2\Real{\Amb^{(\alpha)}_{g\gamma}\overline{B}})\chi_U\rVert_{b/p}
      \label{eq:approxeigen:thm1}
   \end{equation}
   where $1/a+1/b=1$.
   The minimum over $B$ is achieved for $B=\Amb^{(\alpha)}_{g\gamma}$.
\end{mytheorem}
\begin{myproof}
   The proof follows from the middle term of \eqref{eq:lemma:approxeigen:lemmaep3} 
   in Lemma \ref{lemma:approxeigen:lemmaep3} given in the next section
   if one set $W=\BHSpread_{\BH}^{(\alpha)}$ and $K=\chi_U$. The constant
   $\brhoi$ will also be explained later on.
\end{myproof}\\
It follows that for C2, $p=2$ and $a=1$ that 
\begin{equation}
   \eqref{eq:approxeigen:thm1}\leq 2\lVert (1-\Amb_{g\gamma})\chi_U\rVert_\infty
\end{equation}
which improves the previous bounds
\eqref{eq:approxeigen:kozektheorem} and \eqref{eq:approxeigen:kozektheorem:simplified}.
It is independent of the polarization $\alpha$ and does not require any
shape or size constraints on $U$.
Interestingly the offset in \eqref{eq:approxeigen:kozektheorem:simplified},
which does not depend on $(g,\gamma)$
and in a first attempt seems to be related to the notion of underspreadness,
has been disappeared now.

\section{Generalization and Proofs}

For the study of random operators we have
to classify the overall spreading function. 
Thus we assume that all realizations of the
spreading function can be written as
\begin{equation}
   \BHSpread^{(\alpha)}_{\BH}(\mu)=K(\mu)\cdot W(\mu)
\end{equation}
for a common function $K:\Reals^2\rightarrow\RealsPlus$, which 
model some apriori knowledge (for example the
square root of the scattering function in the WSSUS assumption \cite{bello:wssus} 
directly or some
support knowledge). We will always denote with $U\subseteq\Reals^2$ 
then the support of $K$. 
The function $W:\Reals^2\rightarrow\Complexes$ represents
the random part. 
From this considerations it is desirable to measure the error $E_p$ with respect
to a certain $a$--norm $\lVert W\rVert_a$ of the random part, thus to look at the ratio
$E_p/\lVert W\rVert_a$.
We have the following Lemma:
\begin{mylemma}
   Let $\rho_p(\nu):=\lVert\Shift_\nu(\alpha)\gamma-B(\nu)g\rVert_p K(\nu)$.
   For $1\leq p<\infty$, $1\leq a\leq\infty$ and $1/a+1/b=1$ 
   holds 
   \begin{equation}
      E_p/\lVert W\rVert_a\leq\lVert\rho_p\rVert_b
      %<
      %\lVert K\rVert_b\cdot\lVert\gamma\rVert_p+\lVert KB\rVert_b\cdot\lVert g\rVert_p
      \label{eq:approxeigen:lemmaep1:1}
   \end{equation}   
   whenever $W\in\Leb{a}(\Reals^2)$ and $\rho_p\in\Leb{b}(\Reals^2)$.
   \label{lemma:approxeigen:lemmaep1}
\end{mylemma}
\begin{myproof}
   Firstly -- using Weyl's commutation rule and definition of $\lambda$ in 
   \eqref{eq:approxeigen:lambda} gives us
   \begin{equation}
      \begin{split}
         E_p
         &\overset{\eqref{eq:approxeigen:epweyl}}{=}\lVert\int d\nu\,
         \BHSpread_{\BH}^{(\alpha)}(\nu)\Shift_\nu(\alpha)\Shift_\mu\gamma-
         \lambda(\mu)\Shift_\mu g
         \rVert_p \\         
         &\overset{\eqref{eq:weyl:shift:comm}}{=}
         \lVert\Shift_\mu\left( 
         \int d\nu\,\BHSpread_{\BH}^{(\alpha)}(\nu)
         e^{-i2\pi\sympl(\nu,\mu)}
         \Shift_\nu(\alpha)\gamma-
         \lambda(\mu) g\right)         
         \rVert_p \\         
         &\overset{\eqref{eq:approxeigen:lambda}}{=}
         \lVert\int d\nu\, \BHSpread_{\BH}^{(\alpha)}(\nu)
         e^{-i2\pi\sympl(\nu,\mu)}
         (\Shift_\nu(\alpha)\gamma-B(\nu)g)
         \rVert_p\\
      \end{split}
      \label{eq:approxeigen:lemmaep1:proof1}
   \end{equation}
   Note that $p$--norm is with respect to the argument of the functions $g$ 
   and $\Shift_\nu(\alpha)\gamma$.
   The last step follows because $\Shift_\mu(\alpha)$ acts 
isometrically on all $\Leb{p}(\Reals)$.
   If we define 
   \begin{equation}
      f(x,\nu):=e^{-i2\pi\sympl(\nu,\mu)}
      \BHSpread^{(\alpha)}_{\BH}(\nu)[(\Shift_\nu(\alpha)\gamma)(x)-B(\nu)g(x)]
   \end{equation}   
   eq. \eqref{eq:approxeigen:lemmaep1:proof1} reads for $1\leq p<\infty$ by
   Minkowski (triangle) inequality %\cite{lieb:analysis}
   \begin{equation}
      E_p
      =\lVert\int d\nu f(\cdot,\nu)\rVert_p\leq \lVert\int d\nu |f(\cdot,\nu)|\rVert_p
      \leq\int d\nu\lVert f(\cdot,\nu)\rVert_p
   \end{equation}
   %%Equality occurs if...   
   With $\BHSpread_{\BH}^{(\alpha)}=W\cdot K$ and H\"older's inequality 
   follows the claim of this lemma.
\end{myproof}
Let us fix for the moment $\lVert\gamma\rVert_2=\lVert g\rVert_2=1$ 
($\lVert\Amb_{g\gamma}^{(\alpha)}\rVert_\infty\leq 1$) and
a constant $C$ such that $\max(\lVert\gamma\rVert_p,\lVert g\rVert_q)\leq C$. Then
it is obvious that for C1 and C2 follows
$\rho_p(\nu)\leq 2CK(\nu)$ and we have: 
\begin{equation}
   E_p/\lVert W\rVert_a\leq 2C \lVert K\rVert_b
   \label{eq:approxeigen:general:epbound}
\end{equation} 

In the next Lemma we will show that 
$\lVert\rho_p\rVert_b$ can be related to weighted norms of ambiguity function,
which we have studied already in \cite{jung:isit06}. A central role will play 
here the function: 
\begin{equation}
   \brho(\nu):=\sup_x|(\Shift_\nu(\alpha)\gamma)(x)-B(\nu)g(x)|\chi_U(\nu)
   \label{eq:approxeigen:lemmaep3:proof01}
\end{equation}
which characterize the relative smoothness of $g$ and $\gamma$ with respect to
shifts $\nu\in U$. 
Let us furthermore denote its supremum with 
$\brhoi:=\lVert\brho\rVert_\infty$. 
Due to limited space we have to postpone a
detailed discussion of $\brho$ and $\brhoi$ (which will be important for
$p\neq2$ and then one has also to consider $\brhoi=\brhoi(U)$) to a separate journal paper in preparation.

For simplicity we now make \WLOG 
the assumption $\lVert g\rVert_2=\lVert\gamma\rVert_2=1$ and
define the non--negative function $R:=1+|\Amb^{(\alpha)}_{g\gamma}-B|^2-|\Amb^{(\alpha)}_{g\gamma}|^2$. 
\begin{mylemma}
   With the assumptions of Lemma \ref{lemma:approxeigen:lemmaep1}  holds:
   \begin{equation}
      \lVert\rho_p\rVert_b\leq \brhoi^\frac{p-2}{p}\lVert R K^p\rVert_{b/p}^{1/p}
      \label{eq:lemma:approxeigen:lemmaep3}
   \end{equation}
   with equality for $p=2$.
   The minimum over $B$ of the rhs is achieved for C2.    
   \label{lemma:approxeigen:lemmaep3}
\end{mylemma}
\begin{myproof}  
   We have for $p\geq 1$:
   \begin{equation}
      \begin{split}
         \rho_p(\nu)
         &\leq\brho^\frac{p-2}{p}(\nu)\left(
           \int(|(\Shift_\nu(\alpha)\gamma)(x)-B(\nu)g(x)|^2 dx
         \right)^{1/p}\hspace*{-1em}K(\nu)\\
         &=\left(\brho(\nu)K(\nu)\right)^\frac{p-2}{p}\rho_2(\nu)^{2/p}         
      \end{split}
      \label{eq:approxeigen:lemmaep3:proof02}
   \end{equation}
   with equality for $p=2$ such that 
   \begin{equation}
      \lVert\rho_p\rVert_b\leq\brhoi^\frac{p-2}{p}\lVert\rho_2^2K^{p-2}\rVert_{b/p}^{1/p}
      =:\brhoi^\frac{p-2}{p}\lVert R K^p\rVert_{b/p}^{1/p}
      \label{eq:approxeigen:lemmaep3:proof03}
   \end{equation}
   From the definition of $R$ it is obvious that the minimum of the bound in 
   \eqref{eq:approxeigen:lemmaep3:proof02}
   is taken at $B(U)=\Amb^{(\alpha)}_{g\gamma}(U)$ which is provided by C1.
   Because equality for $p=2$ in 
   \eqref{eq:approxeigen:lemmaep3:proof02} this is also the optimizer for 
   $\lVert\rho_2\rVert_b$ for any $b$.  
\end{myproof}
\subsection{Relation to Weighted Ambiguity Norms}
Now we will discuss the connection to weighted norms of ambiguity
functions and fidelity
criteria related to pulse shaping as introduced in \cite{jung:isit06}.
It will give (partially) new insights into the terms of
underspreadness in this context.
Assume that $R_\infty:=\lVert R\chi_U\rVert_\infty\leq 1$ and $b\geq p$. Then it
can be shown that
\begin{equation}
   \begin{split}
      \lVert \rho_p\rVert_b
      &\leq        
      \brhoi^\frac{p-2}{p}\left(\int RK^b\right)^{1/b}
   \end{split}
   \label{eq:approxeigen:lemmaep3:proof0}
\end{equation}
If the latter can not be fulfilled, hence for $b<p$ or if $R_\infty>1$, 
it still holds:
\begin{equation}
   \lVert \rho_p\rVert_b
   \leq
   \brhoi^\frac{p-2}{p} R_\infty^{1/p}\lVert K\rVert_b
   \label{eq:approxeigen:lemmaep3:proof1}
\end{equation}
It can be verified that for C1 the condition $R_\infty\leq 1$ is always fulfilled.
Thus, in this case \eqref{eq:approxeigen:lemmaep3:proof0} reads:
\begin{equation}
   \begin{split}
      \lVert \rho_p\rVert_{b}\leq
      \brhoi^\frac{p-2}{p}\left(\lVert K^b\rVert_1-
        \lVert(\Amb^{(\alpha)}_{g\gamma})^2K^b\rVert_1\right)^{1/b}
      \label{eq:approxeigen:lemmaep3:proof2}
   \end{split}
\end{equation}
We conclude that with our assumptions a maximization of 
the ''2--channel fidelity'' $\lVert(\Amb^{(\alpha)}_{g\gamma})^2K^b\rVert_1$ 
(the case $r=2$ in \cite{jung:isit06}) 
controls $E_p/\lVert W\rVert_a$. This is also the term important for
pulse shaping with respect to scattering function of WSSUS channels 
\cite{kozek:thesis,jung:wssuspulseshaping}.

But for the condition C2 the behavior is different. 
We arrive at the very interesting condition, that $R_\infty\leq 1$
is equivalent to 
\begin{equation}
   \frac{1}{2}\leq\inf_{\mu\in U}\Real{\Amb_{g\gamma}^{(\alpha)}(\mu)}
   \label{eq:approxeigen:lemmaep3:proof3}
\end{equation}
We will show later on that this condition can not be fulfilled on every $U$.
However, if \eqref{eq:approxeigen:lemmaep3:proof3} holds we get
from \eqref{eq:approxeigen:lemmaep3:proof0}
and \eqref{eq:approxeigen:lemmaep3:proof3}:
\begin{equation}
   \begin{split}
      \lVert \rho_p\rVert_{b}
      &\leq
      \brhoi^\frac{p-2}{p}\left(2(\lVert K^b\rVert_1-
        \lVert\Real{\Amb^{(\alpha)}_{g\gamma}}K^b\rVert_1\right)^{1/b} \\
   \end{split}
   \label{eq:approxeigen:lemmaep3:proof4}
\end{equation}
which is then a problem of the maximization of the 
''1--channel fidelity'' (the case $r=1$ in \cite{jung:isit06}).

\section{Some Necessary Support Conditions}
In this section we present some necessary condition on $|U|$
for the case $K=\chi_U$ (therefore support conditions on $\BHSpread_{\BH}^{(\alpha)}$)
which follow from the methods presented in \cite{jung:isit06}.
Due to limited space we have to omit the proofs, which will then appear 
separately in a journal version.
In particular one can show from \cite{jung:isit06} that for 
$|U|\leq e\min(1,2/r)$ follows
\begin{equation} 
   \begin{split}
      \lVert|\Amb_{g\gamma}^{(\alpha)}|^r\chi_U\rVert_1\leq
      |U|
      e^{-\frac{|U|r}{2e}}
   \end{split}
\end{equation}
and with  this result follows:
\begin{mylemma}[Necessary Condition for C2]
   The condition in \eqref{eq:approxeigen:lemmaep3:proof3} can only be fulfilled
   if $|U|\leq 2e\ln 2$.
\end{mylemma}
We believe that this bound is very coarse and it should be 
possible to improve it in using more advanced techniques. 
Further support results follow for $b\geq p$. We define from 
\eqref{eq:approxeigen:lemmaep3:proof2} the following quantity 
\begin{equation}
   r_1(U):= \brhoi^\frac{p-2}{p}\left(|U|-\lVert(\Amb^{(\alpha)}_{g\gamma})^2\chi_U\rVert_1\right)^{1/b}   
   \label{eq:approxeigen:def:r1}
\end{equation}
for the case C1. For C2 we have to guarantee that $R_\infty\leq 1$ and
we define instead from \eqref{eq:approxeigen:lemmaep3:proof4} the quantity:
\begin{equation}
   r_2(U):= \brhoi^\frac{p-2}{p}\left(2(|U|-\lVert\Real{\Amb^{(\alpha)}_{g\gamma}}\chi_U\rVert_1)\right)^{1/b}
   \label{eq:approxeigen:def:r2}
\end{equation}

Then the following can be shown:
\begin{mylemma}[Necessary Conditions on $|U|$]
   Let $1>\delta_k>0$, $|U|\leq e$ and $k=\{1,2\}$.
   If $r_k(U)\leq\delta_k $ then $U$ has to fulfill:
   \begin{equation}
      \brhoi^\frac{p-2}{p}\left(k|U|(1-e^{-\tfrac{|U|}{ek}})\right)^{1/b}\leq\delta_k
      \label{eq:approxeigen:lemmaep4}
   \end{equation}
\end{mylemma}
The latter is an implicit inequality for $|U|$. However, it is possible 
to obtain from this an explicit upper bound for $|U|$ for a given 
$\delta_k$ (not shown in the paper).

%%%%%%%%%%%%%%%%%%%%%%%%%%%%%%%%%%%%%%%%%%%%%%%%%%%%%%%%%%%%%%%%%%%%%%%%%
%%%%%%%%%%%%%%%%%%%%%%%%%%%%%%%%%%%%%%%%%%%%%%%%%%%%%%%%%%%%%%%%%%%%%%%%%
%
\section{Numerical Verification}
In the following we will evaluate and test the obtained bounds 
for Gaussian signaling, i.e. $g$ and $\gamma$ are
time--frequency symmetric Gaussian functions. 
We consider a spreading function given as:
\begin{equation}
   \BHSpread(\nu)=\sum_{k\in\setZ_K^2}c_k\chi_Q(\nu-u(k+o))
   \label{eq:approxeigen:discretespreading}
\end{equation}
where $\setZ_K=\{0\dots K-1\}$, $Q=[0,u]\times[0,u]$ and $o=(\tfrac{1}{2},\tfrac{1}{2})$. 
If we fix the support of the spreading function to be $|U|$, then follows $u=\sqrt{|U|}/K$. 
For such a model the $a$--norm of the spreading function is:
$\lVert\BHSpread\rVert_a=u^{2/a}\lVert c\rVert_a$,
where $\lVert c\rVert_a$ is simply the $a$th vector norm of the vector $c$
with coefficients $c_k$.
\begin{figure}
   \begin{center}
      \includegraphics[width=\linewidth]{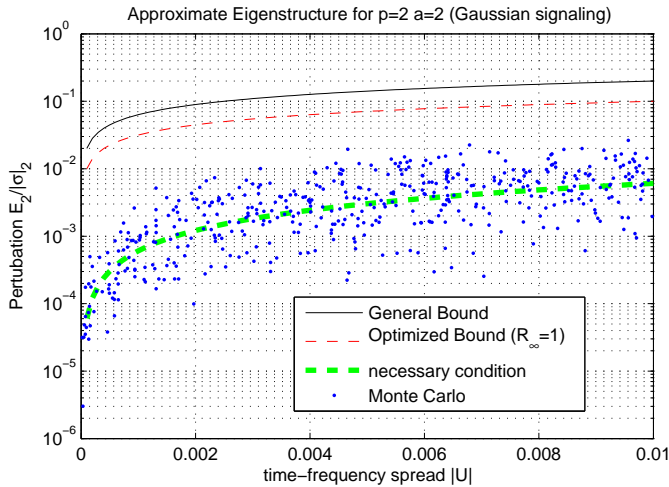}
   \end{center}
   \caption{%
     Approximation error $E_2/\lVert\BHSpread_{\BH}^{(\alpha)}\rVert_2$
     for the case C2. ''General bound'' refers to
     \eqref{eq:approxeigen:general:epbound}, ''Optimized Bound ($R_\infty=1$)'' is
     \eqref{eq:approxeigen:lemmaep3:proof1} and 
     ''necessary condition'' is the results of \eqref{eq:approxeigen:lemmaep4}.
     The monte carlo data is obtained from
     Gaussian signaling and a rectangular spreading function with 
     independent complex normal distributed components $c_k$ ($K=10$) as explained in
     \eqref{eq:approxeigen:discretespreading}.
   }
   \label{fig:approxeigen:B1p2a2}
\end{figure}
The error $E_p$ can be simplified much for Gaussian signaling and finally
computed numerically (therefore Gaussian signaling was chosen). Fig.
\ref{fig:approxeigen:B1p2a2} shows the results of several monte carlo runs,
each corresponds to one point in the plot. For comparison the various bounds 
and results are included in the plot (more details in the caption). Clearly
the most important result (''necessary condition'') can not serve as a bound.
However, its interesting that it produce a rough 
value of the approximation error.

%\begin{figure}
%   \includegraphics[width=\linewidth]{ep_gauss_p=1_a=2.eps}
%\end{figure}
%
%%%%%%%%%%%%%%%%%%%%%%%%%%%%%%%%%%%%%%%%%%%%%%%%%%%%%%%%%%%%%%%%%%%%%%%%%
%%%%%%%%%%%%%%%%%%%%%%%%%%%%%%%%%%%%%%%%%%%%%%%%%%%%%%%%%%%%%%%%%%%%%%%%%

%%%%%%%%%%%%%%%%%%%%%%%%%%%%%%%%%%%%%%%%%%%%%%%%%%%%%%%%%%%%%%%%%%%%%%%%%
%%%%%%%%%%%%%%%%%%%%%%%%%%%%%%%%%%%%%%%%%%%%%%%%%%%%%%%%%%%%%%%%%%%%%%%%%
%
\section{Conclusions}
In this contribution we established in a more general fashion 
the problem of approximate eigenstructure of LTV channels. We extracted
several criteria related to signaling in those channels and pulse shaping.
We hope that our results give some more implications to the role of 
underspreadness for wireless communication. Furthermore the connection 
to symbol calculus of pseudo--differential operators 
is straightforward, such that insights into topics like 
approximate commutativity --- or more generally speaking ---
approximations to spectral properties have to be expected.
%
%%%%%%%%%%%%%%%%%%%%%%%%%%%%%%%%%%%%%%%%%%%%%%%%%%%%%%%%%%%%%%%%%%%%%%%%%
%%%%%%%%%%%%%%%%%%%%%%%%%%%%%%%%%%%%%%%%%%%%%%%%%%%%%%%%%%%%%%%%%%%%%%%%%

\bibliographystyle{IEEEtran}
\bibliography{references}
\end{document}